\documentclass[prd,nofootinbib,twocolumn,showpacs]{revtex4}
\usepackage{graphics}
\usepackage{bm}

\begin{document}

\title{
Reconstruction of field theory from excitation spectra of defects
}

\author{Tanmay Vachaspati}
\affiliation{CERCA, Physics Department, Case Western Reserve University,
10900 Euclid Avenue, Cleveland, OH 44106-7079, USA.}

\begin{abstract}
We show how to reconstruct a field theory from the spectrum of
bound states on a topological defect. We apply our recipe to the case
of kinks in 1+1 dimensions with one or two bound states. Our recipe
successfully yields the sine-Gordon and $\lambda \phi^4$ field
theories when suitable bound state spectra are assumed. The recipe
can also be used to globally reconstruct the inflaton potential of
inflationary cosmology if the inflaton produces a topological defect. 
We discuss how defects can provide ``smoking gun'' evidence for a 
class of inflationary models.
\end{abstract}


\

\maketitle

\section{Introduction}

Often one is interested in diagonalizing a known operator. This is
an eigenvalue problem. The reverse problem, where the eigenvalues
are known but the operator is not, is also of interest.  This is the 
``inverse scattering problem'' \cite{GelLev55,Mar55}  
(also see {\it e.g.} \cite{Solitonbook}). Indeed it
is the basis of the famous question: ``Can one hear the shape of
a drum?'' \cite{Kac66,GorWebWol92}. In other words, can the frequencies 
(eigenvalues) of the sound emitted from the drum be used to determine 
the shape (operator) of the drum?  In the quantum mechanics context,
the equivalent problem is of reconstructing the Schrodinger potential, 
$U(x)$, from the energy spectrum.  Such inverse scattering
problems are of interest in a wide variety of applications. 

We will be interested in a class of problems where $U$ itself originates 
from some field interactions. Our aim is not just to reconstruct the 
Schrodinger potential, $U$, but the underlying {\it field theory} 
interactions that led to that particular Schrodinger potential. 
The eigenvalue data that we will use to reconstruct the field theory
is the excitation spectra of any topological defects that might be 
present in the system. The method is widely applicable since
topological defects are present in a large class of systems. 
They are routinely observed in condensed matter systems, and are 
inevitable in high energy particle physics models. Topological defects 
relevant for high energy particle physics are also relevant for the 
very early universe. Hence the excitation spectra of these topological 
defects can be a window to the early universe, and in particular, to 
inflationary cosmology\footnote{Even if the abundance of topological 
defects is suppressed in the cosmos, the inevitability of defects in 
particle physics implies that they can (in principle) be produced in 
accelerators.}.


A way to understand the usefulness of topological defects in a
reconstruction effort is that the core of a topological defect contains
a different phase of the theory and so the core holds
non-perturbative information about the field theory. 
Therefore the excitation spectrum of the defect leads to a global
reconstruction of the field theory. This is in contrast to a {\it local}
reconstruction that is possible by perturbative methods
such as particle scattering\footnote{In the inflationary case,
astrophysical data can be used with certain assumptions to 
reconstruct the inflaton potential. This reconstruction too
is local since astrophysical observations only probe a very 
limited range of relevant scales.}.

In what follows, we assume that we are given a topological
defect. In fact, for actual calculations, we will exclusively work with 
kinks in 1+1 dimensions since this example illustrates the central 
ideas and also because there are extensive techniques in 1+1 dimensions
that are not available in higher dimensions. (Although spherically
symmetric systems in higher dimensions can be reduced to the 1+1
dimensional case.)

In the next section we will outline the recipe for the reconstruction
and then in Sec.~\ref{examples} explicitly work through two specific 
examples with kinks in 1+1 dimensions. Our scheme yields the sine-Gordon
field theory when there is only one bound state (the translation mode)
on the kink. When the kink has two bound states, we give implicit 
expressions for the field theory. For a specific choice of
the eigenvalues of the two bound states, we obtain the $\lambda \phi^4$ 
field theory. In Sec.~\ref{inflatonpotentials} we discuss the reconstruction 
problem in the inflationary context. Then we qualitatively discuss the bound 
state spectrum on kinks made from the inflaton field. In particular,
we look for signatures of the flatness of the inflaton potential
in the spectrum. Readers only interested in inflation can proceed
directly to Sec.~\ref{inflatonpotentials} since the discussion there is 
largely independent of Sec.~\ref{generalscheme} and \ref{examples}.

\section{General reconstruction scheme}
\label{generalscheme}

We assume that we know the spectrum of energy eigenvalues, 
$\{ \omega_i^2 \}$, for the bound states on a topological defect. 
For the time being, we will only focus on the bound states. The 
price we pay is that there is then a huge degeneracy in the 
reconstruction. By using scattering information and further 
physical input it may be possible to eliminate the degeneracy
but we shall not be discussing this issue here.

The general reconstruction scheme is as follows.

\begin{itemize}
\item[(i)] Prior to an analysis, a theoretical framework is needed
and the appropriate field content should be known.
In this discussion we will only consider a single scalar field
$\phi$ in one dimension, with standard form of the Lagrangian:
\begin{equation}
L = \frac{1}{2} (\partial_\mu \phi )^2 - V(\phi )
\label{lagrangian}
\end{equation}
\item[(ii)] The Schrodinger equation that determines the excitation
spectrum is:
\begin{equation}
\left [ - \frac{d^2}{dx^2} + U(x) \right ] \psi_n = \omega_n^2 \psi_n
\label{schrodinger}
\end{equation}
where
\begin{equation}
U(x) = V'' (\phi_0 (x))
\label{UasVprime2}
\end{equation}
and $\phi_0 (x)$ is the (unknown) profile function of the defect -- 
the classical defect solution. 
We would like to determine the potential $U(x)$. Inverse scattering
methods have been developed precisely to solve this problem.
The answer, however, is not unique, especially if only bound
state spectra are taken into account. However, additional
theoretical input can possibly reduce degeneracies. For example,
if some interactions are known by perturbative methods, the
information might be useful to break some of the degeneracy\footnote{
$U(x)$ may also be reconstructed using scattering data {\it i.e.}
not just the eigenvalues but the scattering amplitudes. For
example, in the Born approximation, the scattering amplitude
is directly related to the Fourier transform of $U(x)$.}.
\item[(iii)] Once we have $U(x)$, we find the ``translation mode'' by 
solving the zero eigenvalue Schrodinger problem:
\begin{equation}
\left [ - \frac{d^2}{dx^2} + U(x) \right ] \psi_t = 0
\label{translationequation}
\end{equation}
\item[(iv)] The translation mode is simply related to the defect profile
functions by differentiation: 
\begin{equation}
\psi_t (x) = \frac{d\phi_0}{dx}
\label{translationmode}
\end{equation}
Hence, we can integrate the translation mode 
to obtain $\phi_0 (x)$.
\item[(v)] Next we invert $\phi_0 (x)$ to obtain $x(\phi_0)$. 
\item[(vi)] The equation of motion for the defect is:
\begin{equation}
- \frac{d^2\phi_0 }{dx^2} + V'(\phi_0 ) = 0
\label{eqofmotion}
\end{equation}
where prime refers to differentiation with respect to the
argument. Combining with Eq.~(\ref{translationmode}) we see that:
\begin{equation}
V'(\phi_0 (x)) = \frac{d\psi_t}{dx} \bigg |_{x(\phi_0)}
\label{Vprime}
\end{equation}
\item[(vii)] Finally, an integration yields the desired symmetry breaking
potential:
\begin{equation}
V(\phi_0) = \int d\phi_0 V'(\phi_0)
\label{desiredV}
\end{equation}
\end{itemize}

In the 1+1 dimensional case described above, there is actually
a slight shortcut that is available to us. In this case, the 
Bogomolnyi equation holds: 
\begin{equation}
\frac{d\phi_0}{dx} = \pm \sqrt{2V(\phi_0)}            
\label{Bogoeq}
\end{equation}
and so:
\begin{equation}
V(\phi_0) = \frac{1}{2} \psi_t^2 \bigg |_{x(\phi_0)}
\label{Bogosoln}
\end{equation}

There are two parts to this recipe. The first is the derivation
of $U(x)$ from the eigenvalue spectrum and the second is the
derivation of $V(\phi )$ from $U(x)$. The part of the recipe 
starting with $U(x)$ and constructing $V(\phi )$ as in 
Eq.~(\ref{Bogosoln}) has also been used in earlier 
work \cite{LimSanRod02}.

The hardest step in this scheme is step (ii), the reconstruction of 
$U(x)$ from $\{ \omega_n^2 \}$. In 1+1 dimensions (or for S-wave states
in 3+1 dimensions), a simple general scheme to reproduce the
bound state spectrum is given in 
Ref.~\cite{KwoRos86,Schetal80,Kwoetal80}. The
scheme employs the idea of supersymmetric quantum mechanics where
the Hamiltonian operator can be factored, and yields a 
{\em reflectionless} potential with the desired bound state spectrum.
We now summarize the scheme; a sketch of how the scheme is 
derived is given in Appendix~\ref{appendixA}. While it may seem
that the reconstruction in terms of supersymmetric potentials is
overly restrictive, we show in Appendix~\ref{appendixB} that 
kinks in one spatial dimension always lead to a supersymmetric
form of the potential $U(x)$. We also show that this is can be
true even when we have multi-component fields in more than one dimension. 
Thus the reconstruction scheme of \cite{KwoRos86} is perfectly suited 
to our context.

Suppose the bound state eigenvalues are labeled in descending order: 
$\omega_i^2 > \omega_{i+1}^2$ and $i=1,...,N-1$ and the zero energy level
is chosen so that the ground state eigenvalue, $\omega_N$, is zero. 
Then a potential containing $n$ of the highest bound states is:
\begin{equation}
U_n (x) = f_n^2 + f_n' + \omega_n^2
\label{Uneqn}
\end{equation}
where the function $f_n (x)$ satisfies:
\begin{equation}
f_n' - f_n^2 + U_{n-1} = \omega_n^2
\label{fneqn}
\end{equation}
in terms of $U_{n-1}$, the potential containing $n-1$ of the
highest bound states.
If we write $f_n(x) \equiv - w_n'/w_n$, the equation for
$w_n$ is:
\begin{equation}
-w_n'' + U_{n-1} w_n = \omega_n^2 w_n
\label{wneqn}
\end{equation}
This equation will have two linearly independent solutions.
If we further require that $U_n$ be even under parity:
$U_n(-x) = +U_n (x)$, then we need $f_n (-x) = - f_n (+x)$,
and $w_n (-x) = w_n (+x)$. This condition eliminates one of
the linearly independent solutions\footnote{Note that the
function $w_n (x)$ is not required to satisfy vanishing boundary
conditions at infinity. In fact, by examining the differential
equation (\ref{wneqn}) in the asymptotic limit, it can be argued 
that $w_n (\pm \infty)$ will be divergent.}.

The solution to the $n^{th}$ Schrodinger equation:
\begin{equation}
\left [ - \frac{d^2}{dx^2} + U_n(x) \right ] \psi_n = \omega_n^2 \psi_n
\label{nthorder}
\end{equation}
is simply:
\begin{equation}
\psi_n = \frac{\alpha_n}{w_n}
\label{psin}
\end{equation}
where $\alpha_n$ is a normalization constant to be determined
by other considerations.
In particular, since $\omega_N$ is the smallest eigenvalue, the
corresponding eigenfunction must be the translation mode of the
defect that does not affect the energy. Therefore $\omega_N =0$
and the translation mode is known once we know $w_N$:
\begin{equation}
\psi_t = \frac{\alpha}{w_N}
\label{psit}
\end{equation}
where we have dropped the subscript on $\alpha_N$.
The profile function is:
\begin{equation}
\phi_0 (x) = \alpha \int \frac{dx}{w_N}
\label{phi0direct}
\end{equation}
and Eq.~(\ref{Bogosoln}) gives:
\begin{equation}
V(\phi_0) = \frac{\alpha^2}{2w_N^2} \biggr |_{x(\phi_0)}
\label{V0direct}
\end{equation} 

The construction of $U_n$ is iterative and one must start 
with the highest bound state with eigenvalue $\omega_1^2$. To find 
$w_1$ we set $U_0 = \omega_0^2$ to be a constant which will be 
determined by other considerations. Now we illustrate this scheme 
in a few cases.

\section{Examples}
\label{examples}

\subsection{One bound state}
\label{oneboundstate}

If the defect has only one bound state, it must be the translation
mode. Since translations do not cost energy, the eigenvalue, $\omega_1$,
vanishes. So we need to solve Eq.~(\ref{wneqn}) with $\omega_1 =0$:
\begin{equation}
-w_1'' + U_0 w_1= 0
\label{onebstateweqn}
\end{equation}
with $U_0 = \omega_0^2$. After some 
manipulations\footnote{We are closely following Ref.~\cite{KwoRos86}.
Note that the eigenvalues in \cite{KwoRos86} are $-\kappa_n^2$ whereas
we have taken them to be $+\omega_n^2$.}:
\begin{equation}
f_1 = -\omega_0 \tanh(\omega_0 x)
\label{onebstatefsolution}
\end{equation}
and
\begin{equation}
U_1 (x) = f_1' + f_1^2 = \omega_0^2 [1-2{\rm sech}^2 (\omega_0 x)]
\label{onebstateU1solution}
\end{equation}

Now we can determine $\omega_0$ by the requirement that at
spatial infinity, the excitations are the particles in the trivial
vacuum. If we denote the masses of these particles by $m$,
this means that
\begin{equation}
U(\infty ) = m^2
\label{Uinfty}
\end{equation}
Hence: $\omega_0 = m$ and
\begin{equation}
U (x) = m^2 [1-2{\rm sech}^2 (m x)]
\label{onebstateUsolution}
\end{equation}

The translation mode, $\psi_t$, satisfies
\begin{equation}
-\psi_t'' + U(x) \psi_t =0
\label{tmodeeqn}
\end{equation}
and, in fact, the solution is
\begin{equation}
\psi_t = \frac{\alpha}{w_1} = \alpha ~{\rm sech} (mx)
\label{psi0soln}
\end{equation}
where $\alpha$ is a constant. Note that, in the present context,
there is no requirement that $\psi_t$ be normalized as a wavefunction.

Therefore the profile function is
\begin{equation}
\phi_0 (x) = \alpha \int dx ~ {\rm sech} (mx)
= \frac{2\alpha}{m} 
\tan^{-1} \biggl [{\rm tanh} \biggl (\frac{mx}{2} \biggr ) \biggr]
\label{phi0solution}
\end{equation}
In other words,
\begin{equation}
{\rm tanh} \biggl ( \frac{mx}{2} \biggr ) = 
\tan \biggl ( \frac{m\phi_0}{2\alpha} \biggr )
\label{xphi0relation}
\end{equation}

Since here the Bogomolnyi equations can be used, Eq.~(\ref{Bogosoln}) 
with some algebra gives:
\begin{equation}
V(\phi_0 ) = \frac{\alpha^2}{4}\biggl [ 
              \cos \biggl ( \frac{2m\phi_0}{\alpha} \biggr ) +1
                              \biggr ]
\label{Vofphi0}
\end{equation}
This is the sine-Gordon potential.

The potential still contains the unknown parameter $\alpha$.
To fix this parameter we could use some other property of the
defect, for example, its total energy. 

This completes the global reconstruction of the potential in the 
single bound state case.

\subsection{Two bound states}
\label{twoboundstates}

Now there are two eigenvalues $\omega_1^2$ and $\omega_2^2$.
The translation mode is always the lowest eigenvalue and hence
$\omega_2^2=0$. So we first need to find a potential, $U_1(x)$ that 
contains the $\omega_1^2$ mode. 

Following the recipe given in the previous section, we have
\begin{equation}
U_1 = f_1^2+f_1'+\omega_1^2
\label{u1formula}
\end{equation}
and we need to solve
\begin{equation}
-w_1''+\omega_0^2 w_1 = \omega_1^2 w_1
\end{equation}
where, as before $U_0 = \omega_0^2$. This is exactly the single
bound state problem that we solved in the previous subsection if
we replace $\omega_0^2$ with $\nu^2 \equiv \omega_0^2 -\omega_1^2$. 
Therefore,
\begin{equation}
f_1 = - \nu \tanh (\nu x)
\end{equation}
and
\begin{equation}
U_1 = \nu^2 [ 1-2 {\rm sech}^2 (\nu x) ] + \omega_1^2
\label{u1for2bstate}
\end{equation}

Now we use the second eigenvalue. The potential $U_2$ is given by
\begin{equation}
U_2 = f_2^2+f_2'+\omega_2^2 = f_2^2 + f_2'
\label{u2formula}
\end{equation}
where we have used $\omega_2^2=0$ since this mode must be the
translation mode. So we need to solve:
\begin{equation}
-w_2''+  
  \{ \nu^2 [ 1-2 {\rm sech}^2 (\nu x) ] + \omega_1^2 \} w_2 = 0
\label{w2equation}
\end{equation}
We can rescale $z=\nu x$ and bring the equation to the form:
\begin{equation}
\frac{d^2w_2}{dz^2} +  
\{ \lambda + 2 {\rm sech}^2 (z) \} w_2 = 0
\label{neww2equation}
\end{equation}
where:
\begin{equation}
\lambda \equiv - \left [ 1 + \frac{\omega_1^2}{\nu^2} \right ]
\end{equation}
Eq. (\ref{neww2equation}) has been solved in Ref.~\cite{MorFes53}
(see Section 6.3, page 768) and the solution is given in terms of
hypergeometric functions:
\begin{eqnarray}
w_2 &=& \{ \xi (1-\xi ) \}^{K/2} \biggl [ 
    \vphantom{F}_{2}F_{1}(K+2, K-1; K+1; \xi ) \nonumber \\
  &+&   \vphantom{F}_{2}F_{1}(K+2, K-1; K+1; 1-\xi ) \biggr ]
\label{w2solution}
\end{eqnarray}
where,
\begin{equation}
\xi \equiv \frac{1+\tanh(z)}{2}
\end{equation}
and
\begin{equation}
K^2 \equiv - \lambda
\end{equation}
Note that we have fixed the ratio of the two linearly independent
solutions in Eq.~(\ref{w2solution}) by imposing the requirement
that $w_2$ have even parity under $z \rightarrow -z$, which is the 
same as requiring $w_2(\xi ) = w_2 (1-\xi )$. (The overall normalization 
of $w_2$ is unimportant.) Now Eq.~(\ref{V0direct}) immediately gives 
the field theoretic potential though the expression is still implicit 
since the inverse function, $x(\phi_0)$, needs to be determined after 
doing the integral in Eq.~(\ref{phi0direct}).

Let us now look at the special case when $\omega_1^2/\nu^2 = 3$
($K = 2$). Then the solution for $w_2$ is given by elementary 
functions:
\begin{equation}
w_2 (x) = \cosh^2 (\nu x) 
\label{w2special}
\end{equation}
With this 
\begin{equation}
\phi_0 (x) = \alpha \nu \int \frac{dx}{w_2}
           = \alpha \tanh (\nu x)
\end{equation}
where $\alpha$ is a constant of normalization. (For convenience
we have explicitly included a factor of $\nu$ in the normalization.)
Therefore:
\begin{equation}
V(\phi_0 ) = \frac{\alpha^2}{2} {\rm sech}^4 (z) 
           =  \frac{\alpha^2}{2} 
               \left ( 1 - \frac{\phi_0^2}{\alpha^2} \right )^2
\label{V0special}
\end{equation}
Hence we have recovered the $\lambda \phi^4$ model in this special case.

It appears remarkable that the simplest examples of one and two
bound states have led to two well-known field theories, namely the
sine-Gordon and $\lambda \phi^4$ theories. A partial explanation
is to be found in section \ref{appendixB}, where we show that
whenever $U(x)$ arises from a field theoretic kink, it must
necessarily be of the supersymmetric form (Eq.~(\ref{Uneqn})).
So the recipe to reconstruct $U(x)$ is ideally suited to the present
problem.

\subsection{Three or more bound states; higher dimensions}
\label{threeormore}

The reconstruction can be continued to the case of three of more 
bound states. However, the general reconstruction becomes difficult
to do analytically and numerical methods are desirable. Such
methods have not yet been developed.

The inverse scattering technique that we have used only works in
one spatial dimension. Problems in more than one spatial dimension 
require other inverse scattering techniques. However, if the
problem is spherically symmetric, the problem effectively reduces
to one spatial dimension and the reconstruction outlined above
can be applied. (In Ref.~\cite{KwoRos86} the authors applied
the inverse scattering technique to S-wave bound states of quarks.)

\section{Application to cosmological inflation}
\label{inflatonpotentials}

\subsection{Inflation and defects}
\label{inflation}

If there was an extended but limited period of superluminal expansion
in the early universe, light scalar fields would get excited by the
spacetime expansion and
create adiabatic perturbations on superhorizon scales. The prediction
of adiabatic perturbations matches observations of anisotropies in the 
Cosmic Microwave Background Radiation (CMBR) and is the reason for the 
current confidence
in the cosmic ``inflationary'' scenario. The mechanism driving 
cosmic inflation is a scalar field called the ``inflaton''. The
dynamics of both the spacetime and the scalar field is dominated by the
inflaton potential energy for the entire duration of cosmic inflation. 
Under certain assumptions, this requires the scalar field to lie on 
top of a very flat
potential. With time the scalar field slowly rolls along the flat
potential, eventually reaching the steeper parts of the potential
where its kinetic energy becomes significant. Then inflation stops
and the subsequent evolution of the scalar field leads to ``reheating''
{\it i.e.} conversion of scalar field kinetic energy into ordinary matter
in a thermal state\footnote{There are models of inflation that do not
require a flat potential. Instead they depend on a large initial value 
of the scalar field. Here we are limited to inflationary models
that rely on the potential being very flat.}.

One hopes that the scalar fields necessary for cosmic inflation 
will automatically arise in high energy particle physics models, 
such as Grand Unified Theories (GUTs) or string theory. On very
general grounds, it is known that such theories contain magnetic
monopoles and sometimes other topological defects. 
If inflation and the defect are due to separate fields, the study 
of the defect core cannot yield information about inflation.
For the global reconstruction scheme under discussion to work,
the inflaton itself should play a role in the structure 
of the topological defect. 

One of the cosmological problems that inflation was designed to
solve was the magnetic monopole overabundance problem that arose
out of the marriage of GUTs and cosmology. If there is a period
of inflation during or after the GUT phase transition, the magnetic 
monopole density would get diluted to insignificant amounts, leaving
perhaps only a few in the entire visible universe. However, subsequent 
work on inflation has shown that topological defects can still be
produced in significant numbers toward the late stages
\cite{KofLinSta96,Tkaetal98,KasKaw98,Feletal00,RajCop00}. 
If bound states on one of these defects can be studied, it would help 
in the global reconstruction of the inflaton potential\footnote{Often
the defects are due to topology present due to fields other than
the inflaton. This significantly complicates the reconstruction
though, in principle, if there is variation of the inflaton within
the defect, a reconstruction would still be possible.}. The other 
possibility of using 
the reconstruction scheme is if future particle physics experiments 
are able to produce magnetic monopole and antimonopole
pairs (or closed walls or strings) in the laboratory\footnote{
If inflation occurred at a relatively low energy scale -- say
somewhat larger than the electroweak scale -- this possibility is 
easier to envision and might even be realized at energies available 
at the Large Hadron Collider (LHC). However, the production of
solitons by scattering particles is expected to be difficult
because a large number of particles are simultaneously involved.}. 
The excitations of these monopoles could then be studied. There is 
also the possibility that duality holds in particle physics and the 
known particles ({\it e.g.} the electron) may be dualized magnetic 
monopoles \cite{Vac96}. The internal excitations of the electron, 
would then be excitations of a magnetic monopole. Perhaps these 
excitations can tell us something about inflation. 

Earlier work on reconstructing the inflaton potential 
(see {\it e.g.} \cite{Lidetal97}) worked under the assumption
that the CMBR anisotropies are generated by quantum fluctuations 
of the inflaton. This results in a {\it local}
reconstruction of the potential, the limited range of scales
observed being explained entirely by a very small portion of the 
entire inflaton potential. Here we will consider signatures of
{\it global} features of the inflaton potential, such as flatness.

Before we proceed to discuss signatures of inflaton potentials in
the excitation spectrum, it is worth noting that inflation can occur
within topological defects, so-called ``topological inflation''
\cite{Vil94,Lin94}. We will, however, not consider this
possibility since, if a defect inflates, it is not possible to
find it or create it within our horizon while retaining 
predictability \cite{BorTroVac99}. The condition for the topological
defect to start inflating is that its width, $\delta$, be larger
than the horizon size, $1/H$, corresponding to the energy density
within the defect.  As we will show below
(Eq.~(\ref{deltavalue})), in the case of domain walls with
very flat potentials, $\delta \sim \eta/\sqrt{2V(0)}$, where
$\eta$ is the change in $\phi$ across the wall. 
The cosmological equations give $H^2 = 8\pi G V(0)/3$.
Hence the condition for topological inflation is 
$\eta > m_{P}$ where $\eta$ is the vacuum expectation value of the 
field and $m_P$ is the Planck mass. Hence we will be restricted to 
$\eta \ll m_P$ and we will ignore gravitational effects.

\subsection{Properties of the excitation spectrum}
\label{properties}

The reconstruction of the inflaton potential can be carried out following 
the recipe given in Sec.~\ref{generalscheme} if the kink bound 
state spectrum is known. Here we will consider what is essentially
the scattering problem for the inflaton kink, namely the question: 
what signatures might we see in the bound state spectrum if the kink 
on hand is due to the inflaton? 

Consider a scalar field, $\phi$, with potential $V(\phi )$ that is
invariant under the $Z_2$ transformation $\phi \rightarrow -\phi$ 
{\it i.e.} $V(-\phi )=V(+\phi )$. Let the true vacua be given by
$\phi =\pm\eta$. Then there will be a domain wall solution across 
which $\phi$ will change from $-\eta$ to $+\eta$. We are interested
in the case when $\phi$ is also an inflaton with very flat $V(\phi)$ 
for $\phi \in (-\eta, +\eta)$. Here we will show that the flatness 
of the inflaton potential $V(\phi)$ has definite predictions
for the spectrum of excitations of the domain wall\footnote{In 
non-Abelian field theories, domain wall solutions are much more 
complex. For example, domain walls in $SU(N)\times Z_2$ have been 
discussed in Ref.~\cite{PogVac00,Vac01}. We will only discuss the 
simplest case of a $Z_2$ kink with a single scalar field in this 
paper.}. 

\begin{figure}
\scalebox{0.40}{\includegraphics{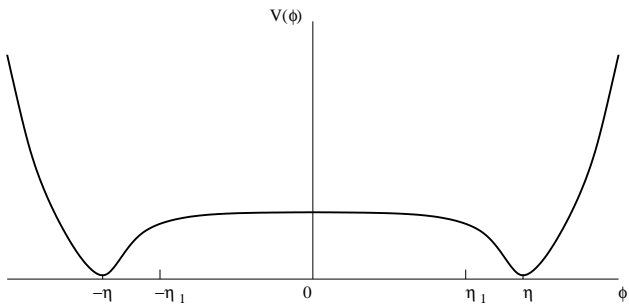}}
\caption{Sketch of an inflaton potential. A crucial feature is
the broad flat region from $-\eta_1$ to $+\eta_1$. The width
of the curved region from $\pm \eta_1$ to $\pm \eta$ is assumed 
to be much smaller than $\eta_1$.
}
\label{flatpot}
\end{figure}

\begin{figure}
\scalebox{0.40}{\includegraphics{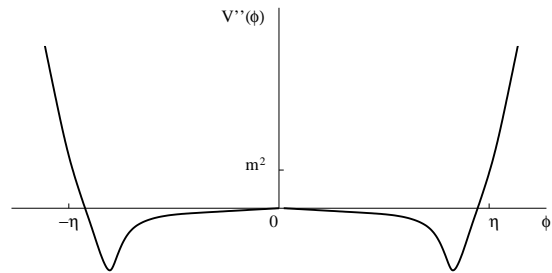}}
\caption{Sketch of the second derivative of the inflaton potential. 
}
\label{V2deriv}
\end{figure}

\begin{figure}
\scalebox{0.40}{\includegraphics{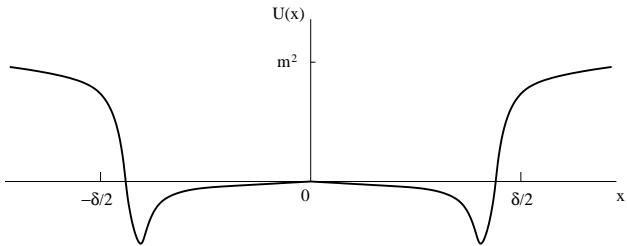}}
\caption{Sketch of $U(x)$.
}
\label{Usketch}
\end{figure}

A flat potential is drawn in Fig.~\ref{flatpot}. In the
discussion below we will assume that $(\eta - \eta_1) \ll \eta_1$,
so that the transition from flat to curved potential occurs
relatively quickly\footnote{If this assumption is not valid,
the discussion below of the bound state spectrum will need
to be modified.}.  A rough approximation to the potential in the 
interval $(-\eta,+\eta )$ is 
given by a top hat with some modification near the curved regions 
in the vicinity of $\phi = \pm \eta_1$. The shape of the potential 
for $\phi > \eta$ will not be crucial for us but we will use the
fact that the curvature at the global minimum is $+m^2$ where
$m$ is the mass of small excitations infinitely far away from
the domain wall. The second derivative of the potential $V''(\phi )$ 
is sketched in Fig.~\ref{V2deriv}. Note that the horizontal axis is 
$\phi$ in this plot. The potential $U(x)$ that determines the bound 
states is given by:
\begin{equation}
U(x) = V'' (\phi_0 (x))
\label{UVdprime}
\end{equation}
where $\phi_0 (x)$ is the domain wall solution. (The problem
is effectively 1+1 dimensional and hence we suppress dependencies
on the $y$ and $z$ coordinates.) For the top hat potential:
\begin{equation}
\phi_0 (x) = \left \{
\begin{array}{ll}
 -\eta & \mbox{$x < -\delta/2$} \\
\eta (2x/\delta ) & \mbox{$-\delta /2 \le x \le \delta/2 $} \\
+\eta & \mbox{$x > +\delta/2$} 
\end{array}
            \right.
\label{phi0tophat}
\end{equation}
where $\delta$ is the thickness of the defect and is determined
by the Bogomolnyi equation (\ref{Bogoeq}) as:
\begin{equation}
\delta = \eta \sqrt{\frac{2}{V(0)}}
\label{deltavalue}
\end{equation}
Therefore the potential $U(x)$ has the shape shown in 
Fig.~\ref{Usketch}. The broad flat bottom of the potential in 
the central region and the asymptotic behavior are generic to the 
inflationary models we are considering. The details of the transition 
regions near $|x| \sim \delta/2$ may be model-dependent since they 
depend on the transition from flat $V(\phi )$ where the field rolls 
slowly to the curved part of $V(\phi )$ where reheating starts. 
The general features of $U(x)$ are that is has a very broad central 
region where $U(x) \sim 0$, then a dip, then a rise that occurs in 
a width that is much smaller than $\delta$, and finally an asymptotic 
plateau of $m^2$.

We know that the lowest bound state is the translation mode and 
has vanishing energy. In fact, Eq.~(\ref{phi0tophat}) can be
used to find the translation mode $\psi_t$ for the top hat case:
\begin{eqnarray}
\psi_t (x) = \frac{d\phi_0}{dx} = \left\{ 
\begin{array}{ll}
{2\eta}/{\delta} &  \mbox{$-\delta /2 \le x \le \delta/2$} \nonumber \\
    0 & {\rm otherwise}
\end{array}
             \right.
\label{psi0tophat}
\end{eqnarray}
Since the translation mode is the lowest energy bound state
 -- an eigenstate with lower eigenvalue would signal an instability 
of the defect --  all bound states have positive energy and lie above 
the top of the double well structure in the transition regions. 
So we expect the higher bound states to be relatively insensitive
to the details of the transition region and mimicking $U(x)$ 
by a finite square well potential may be a reasonable way to
start a first analysis.

The finite square well potential is analyzed in virtually every
quantum mechanics textbook ({\it e.g.} \cite{Gri95}). If $U_\infty$ 
denotes the depth of the square well and $\delta$ its width, the 
number of bound states of a particle of mass $\mu$ is given by:
\begin{equation}
N_b \sim \sqrt{\mu U_\infty \delta^2}
\label{Nbndstates}
\end{equation}
In our case (see Eq.~(\ref{schrodinger})), $2\mu =1$, $U_\infty = m^2$, 
and $\delta$ is given by Eq.~(\ref{deltavalue}). Therefore:
\begin{equation}
N_b \sim \frac{m\eta}{\sqrt{V(0)}}
\label{Nourcase}
\end{equation} 

The three parameters $m$, $\eta$ and $V(0)$ occurring in this
formula are independent: $m^2$ is the curvature in the true vacuum,
$V''(\eta )$; $\eta$ is the position of the true vacuum; $V(0)$
is the height of the flat part of the potential. If all these 
parameters were constrained by inflationary cosmology, we would 
have some bounds on $N_b$. However, there are no useful constraints 
on the parameter $m$ and no model-independent bounds on $N_b$ can
be derived. So, to get an idea of the range of $N_b$, we work
out its value in the case when $V(\phi )$ is of the Coleman-Weinberg 
form \cite{ColWei73} with vanishing curvature at $\phi =0$
\cite{KolTur94}:
\begin{equation}
V(\phi ) = \frac{B}{2}\sigma^4 + 
 B \phi^4 \left [ 
     \ln \left ( \frac{\phi^2}{\sigma^2} \right ) - \frac{1}{2} 
          \right ] 
\label{colemanweinberg}
\end{equation}
We then find:
\begin{equation}
N_b \sim 5
\label{nbcolemanweinberg}
\end{equation}
Hence we expect that an inflaton domain wall will have at least
a few bound states.

As is well-known \cite{Gri95}, a special property of the infinite 
square well potential is that its eigenvalue spectrum is proportional 
to $n^2$ where $n=1,2,...$ is an integer that labels the eigenstates 
starting with the ground state. The same property holds for the low 
lying eigenstates of the finite square well\footnote{The exact 
eigenvalues depend on solutions of transcendental equations and must 
be found numerically.}. If the potential is shaped like 
$\sim |x|^b$, then it can be shown that $E_n \propto n^{2b/(b+2)}$ 
(see Appendix~\ref{appendixC}). This dependence 
shows that observations of the spectrum of bound states on a domain 
wall can be used to find the power $b$ and hence the shape of the 
potential. In the inflationary case, the width of the well is
much larger than the distance over which the sides of the well
get to their asymptotic levels. Therefore, the sides of the well
are very steep ($b \gg 1$) in relation to the width of the wall
and so the bound state spectrum $\omega_n^2$ should be proportional 
to $\sim n^2$.

Qualitatively similar arguments may be used in the three dimensional 
case, in the case when the bound state spectrum on a monopole is known. 
If the potential, $U({\bf x})$, is spherically
symmetric, the density of energy eigenstates with fixed total
angular momentum can once again lead to information about the 
flatness of the inflaton potential. For example, consider
the s-wave states in a spherically symmetric potential
$U(r)$. The Schrodinger equation for the $n^{th}$ eigenstate
radial wave function $R_n (r)=u_n(r)/r$ reduces to:
\begin{equation}
-\frac{1}{2\mu} \frac{d^2 u_n}{dr^2} + U(r) u_n
 = E_{n0} u_n
\label{radialeq}
\end{equation}
where the energy eigenvalue $E_{n0}$ carries the $l=0$ label.
As in the one dimensional case, here too we expect 
$E_{n0} \propto n^2$ when $U(r)$ is due to an inflaton.

\section{Conclusions}
\label{conclusions}

We have described a ``recipe'' for recovering the potential
in a field theory, $V(\phi )$, starting with the bound state
spectrum on a topological defect. An important aspect of the
reconstruction discussed here is that it is ``global'' -- the
whole potential is reconstructed and not just a small part of
it. As specific examples, we have applied the recipe to the
case of one and two bound states on kinks. In the one bound state 
case, the recipe yields the sine-Gordon potential and in the two
bound state case with a specific set of eigenvalues we obtained
the $\lambda \phi^4$ potential.

While implementing the recipe we have relied on the inverse 
scattering method based on supersymmetric quantum mechanics 
\cite{KwoRos86}. The method yields reflectionless potentials $U(x)$.
Other inverse scattering methods may also be used and, in general,
they will lead to different $U(x)$ with the same bound state
spectra. The non-uniqueness of the reconstruction may be
reduced by further inclusion of scattering data and perhaps by
using other system-specific information.

Our recipe for the reconstruction of $V(\phi )$ is sufficiently
involved that, except in the simplest situations, it will have to 
be implemented numerically. For example, even in the two bound state 
case, the solutions of the differential equations in the inverse
scattering problem are given in terms of hypergeometric functions,
making it very hard to analytically implement the recipe.
The reconstruction in 3+1 dimensions is technically even more
challenging. However, we can expect the analysis to be close
to the 1+1 dimensional case when the problem is spherically
symmetric. 

The results of this paper are relevant to any system in which
topological defects occur. Hence we can expect applications to
condensed matter systems, particle physics, and cosmology. It
would be interesting to try out the reconstruction in condensed
matter systems where topological defects are readily available.
(Or in nuclear physics to the extent that nuclei can be modeled
by Skyrmions \cite{Sky61,AdkNapWit83,skyrmereviews}.) 
The reconstruction would yield a Landau-Ginzburg 
type of effective interaction potential but would not yield, at 
least directly, information about the microphysical interactions 
between the fermions. The application to particle physics and 
cosmology is futuristic since topological defects are theoretically 
expected in these settings but have not yet been experimentally 
discovered or observed. Just as in the condensed matter case, the 
scalar field need not be fundamental even in the particle physics 
context. 

A novel application of the reconstruction recipe is in the context 
of inflationary cosmology. If the inflaton vacuum manifold has 
suitable topology, topological defects in the inflaton field will 
exist. The spectrum of bound states in these defects will reflect the
properties of the inflationary potential. Reversing the argument,
since we know that the inflationary potential must have certain
properties, the bound state spectrum must also have some
characteristics. We have discussed these characteristics as a
way to probe the inflationary scenario, which is hard to do 
otherwise \cite{LueStaVac03}. If future investigations discover
a scalar field with an extremely flat potential, that scalar field 
will be a prime suspect to be the inflaton, and the defect with
its characteristic bound state spectrum will be a ``smoking gun'' 
from the shot that was fired ten billion years ago.

If the topology of the inflaton vacuum manifold is trivial, no 
defects will exist and the reconstruction scheme discussed here 
will not be useful for inflationary cosmology. If, however, inflaton 
topological defects do exist, experiments may become feasible in 
the future that can directly probe cosmology in the laboratory.

\begin{acknowledgments} 
These ideas were inspired during the ``Cosmology in the Laboratory''
(COSLAB 2003) meeting, Bilbao. I am grateful to Carl Bender, Dorje 
Brody and Jon Rosner for pointing out references to work on the inverse
scattering problem, to Grisha Volovik for comments on the situation
in condensed matter systems, to Martin Bucher, Jaume Garriga and
Alex Vilenkin for discussions about cosmological applications, and 
to Craig Copi and Harsh Mathur for useful suggestions. I acknowledge
hospitality at Imperial College, University of Barcelona, and DAMTP
(University of Cambridge) while this work was being done. This work 
was supported by DOE grant number DEFG0295ER40898 at Case.

\end{acknowledgments}

\appendix

\section{Check of the inverse scattering equations}
\label{appendixA}

Here we give a check of the iterative scheme for the
inverse scattering method described in Ref.~\cite{KwoRos86}.

Suppose we are given a potential $U_{n-1}(x)$ that contains
$n-1$ of the highest bound states. Then we construct the 
functions $f_n(x)$ from the equation:
\begin{equation}
f_n' - f_n^2 +U_{n-1} = \omega_n^2
\end{equation}
This non-linear equation is simplified by setting
\begin{equation}
f_n = -\frac{w_n'}{w_n}
\end{equation}
using which we get the linear equation
\begin{equation}
- w_n'' + U_{n-1} w_n = \omega_n^2
\end{equation}
Finally we construct
\begin{equation}
U_n (x) = f_n' + f_n^2 + \omega_n^2
\end{equation}
The claim is that $U_n$ has an eigenstate with eigenvalue
$\omega_n^2$ in addition to all the other higher eigenstates.

That $U_n$ admits a state with eigenvalue $\omega_n^2$, can be
shown explicitly. The state is given by:
\begin{equation}
\psi_n = \frac{\alpha}{w_n}
\end{equation}
Then it easy to check that
\begin{equation}
-\psi_n '' + U_n \psi_n = \omega_n^2 \psi_n
\end{equation}
So, indeed, the potential $U_n$ has an (explicitly constructed) 
eigenstate with eigenvalue $\omega_n^2$.

For the iterative procedure to work, we also need to show
that $U_n$ admits eigenstates with the higher eigenvalues 
$\{ \omega_{i}^2 \}$ for $i=1,...,n-1$. This is seen as follows.
$U_n$ and $U_{n-1}$ are ``partner'' potentials. In other words,
we can write the $n^{th}$ Hamiltonian as:
\begin{equation}
H = (-\partial_x^2 + U_n ) = A^+ A + \omega_n^2
\end{equation}
where
\begin{equation}
A = - \partial_x + f_n \ , \ \ 
A^+ = + \partial_x + f_n 
\end{equation}
Then, the partner Hamiltonian is:
\begin{equation}
H_- \equiv (-\partial_x^2 + U_{n-1} ) = A A^+ + \omega_n^2
\end{equation}
It is now easy to show that if $\psi$ satisfies
\begin{equation}
H\psi = E\psi
\end{equation}
then $A\psi$ is an eigenstate of $H_-$ with the same eigenvalue:
\begin{equation}
H_- (A \psi) = E (A \psi)
\end{equation}
Thus, $H$ and $H_-$ have a common eigenspectrum, except for the
lowest state of $H$ that satisfies $A\psi =0$. Hence $U_{n}$ has
states with eigenvalues $\omega_i^2$, $i=1,...,n-1$ since these
are also the eigenvalues of states in $U_{n-1}$, and then it has
one extra eigenstate and this has eigenvalue $\omega_n^2$.

\begin{figure}
\scalebox{0.40}{\includegraphics{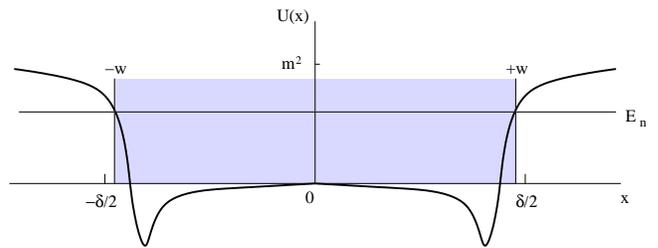}}
\caption{Sketch of $U(x)$ together with energy level $E_n$.
The shaded region is the interior of the square well 
potential, $U_{sq}$, that is used to find the approximate 
dependence of $E_n$ on $n$.
}
\label{appfig}
\end{figure}

\section{Form of domain wall fluctuation potentials}
\label{appendixB}

Here we show all that all kink potentials ($U$) have the
supersymmetric form. We know that every kink has a 
translation mode which does not change the energy. Therefore, the
Schrodinger equation (Eq. ((\ref{schrodinger})) gives
\begin{equation}
U(x) = \frac{\psi_t ''}{\psi_t}
\end{equation}
where $\psi_t (x)$ is the translation mode. This form
of $U(x)$ can be rewritten as:
\begin{equation}
U(x) = f'+f^2
\end{equation}
where $f = [\ln(\psi_t)]'$, and hence $U$ is of the supersymmetric
form (Eq.~(\ref{Uneqn})) \cite{RodFilVai98}.

This argument can also be extended to higher dimensions with
derivatives replaced by higher dimensional derivatives. Then
we find:
\begin{equation}
U = {\bf \nabla} \cdot {\bf f} + {\bf f}^2
\end{equation}
where ${\bf f}({\bf x}) = {\bf \nabla} (\ln \psi_t )$ is a
vector valued function and $\psi_t ({\bf x}$ is a translation
mode. The Hamiltonian is:
\begin{equation}
H = {\bf A}^+ \cdot {\bf A}
\end{equation}
where
\begin{equation}
{\bf A} = -{\bf \partial} + {\bf f}
\end{equation}

If $\psi$ is a multi-component scalar field, or a collection
of several scalar and gauge field fluctuations, in several 
dimensions, there are circumstances in which $U$ is still
supersymmetric. Label the many components of $\psi$ by the
index $i$. So $\psi$ may be thought of as a column vector
with components $\psi_i$. Now in higher dimensions, there will
be several zero modes. For example, translations along any
dimension will be a zero mode. So label the zero modes by
the index $a$ and denote the translation modes by $\xi^a$.
Then consider the matrix ${\bf M}$ whose components are
$\xi^a_i$. We can show that $U$ is supersymmetric if we
assume that ${\bf M}$ is a square matrix that is invertible.
This follows because now:
\begin{equation}
[-{\bf \nabla}^2 + U({\bf x}) ] {\bf M} = 0
\label{schrodM}
\end{equation}
($U$ itself is a matrix potential.) Then it is easy to show 
that
\begin{equation}
U({\bf x}) = {\bf \nabla}\cdot {\bf F} + {\bf F}^2
\label{Umulticomponent}
\end{equation}
where
\begin{equation}
{\bf F } = ({\bf \nabla} {\bf M}) {\bf M}^{-1}
\label{bfF}
\end{equation}

\section{Spectral properties of bound states in 1+1 dimensions}
\label{appendixC}

We now find the connection between the shape of the 
potential and the dependence of the $n^{th}$
eigenvalue, $E_n$, on $n$ for the specific class of
potentials:
\begin{equation}
U(x) = a |x|^b
\label{genU}
\end{equation}
where $a$ and $b$ are parameters. 

We use the WKB approximation \cite{Gri95}. The quantization
condition for a particle of mass $\mu =1/2$ in the potential
$U(x)$ is
\begin{equation}
\int_{-w(E_n)}^{+w(E_n)} dx \sqrt{E_n - U(x)} = n
\end{equation}
where $E_n$ stands for $\omega_n^2$ in the notation of
Sec.~\ref{generalscheme}.  Inserting the class of potentials 
in Eq.~(\ref{genU}), we find:
\begin{equation}
E_n \propto n^{2b/(b+2)}
\end{equation}

Next we conside the WKB method in the context of the potential 
$U(x)$ redrawn here in Fig.~\ref{appfig}.

Now,
\begin{eqnarray}
n &=& \int_{-w(E_n)}^{+w(E_n)} dx \sqrt{E_n - U(x)} \nonumber \\
  &=& 
        \int_{-w(E_n)}^{+w(E_n)} dx 
               \sqrt{(E_n -U_{sq}) - (U-U_{sq})} 
\label{intsplit}
\end{eqnarray}
$U_{sq}$ denotes the infinite square well potential of width
$2w(E_n)$ (see Fig.~\ref{appfig}). Our assumption is that
the dominant contribution to the integral comes from the
$E_n-U_{sq}$ term and that the $U(x)-U_{sq}$ term can be
ignored. Then the integral is trivial to do, resulting in
$n = \sqrt{E_n} 2 w(E_n)$. Now $2 w(E_n) \sim \delta$.
Therefore $E_n \sim n^2$.

If $E_n$ is very small, our assumption that $U(x)-U_{sq}$
can be ignored will not hold. For reasonably large $E_n$ 
we can expect it to hold, though $E_n$ should still be
much smaller than $m^2$ -- the asymptotic value of $U$ -- so 
that the finite depth of the well does not play a role.

\end{document}